# Scalable Federated Learning over Passive Optical Networks

Jun Li[1], Lei Chen[2], and Jiajia Chen[1]

[1]Department of Electronic Engineering, Chalmers University of Technology, Göteborg, Sweden
[2]Research Institutes of Sweden, Göteborg, Sweden.
{ljun, jiajiac}@chalmers.se

**Abstract:** Two-step aggregation is introduced to facilitate scalable federated learning (SFL) over passive optical networks (PONs). Results reveal that the SFL keeps the required PON upstream bandwidth constant regardless of the number of involved clients, while bringing ~10% learning accuracy improvement.

## 1. Introduction

Federated learning (FL) is a privacy-preserving paradigm of distributed learning, where clients (e.g., mobile terminals, user devices) can collaboratively learn a shared model while keeping all the data locally [1]. During the training process of FL, each client needs to periodically transmit its local model parameters to the centralized parameter server (CPS), where a set of global model parameters are updated according to aggregation strategies such as the federated averaging algorithm (FedAvg) [1]. Then, the CPS sends the global model parameters to each client to update its local model for a new round of training. One challenge here is that the amount of training traffic generated by the model updates can be huge, particularly for the upstream which often has limited bandwidth compared to the downstream. For example, the training traffic for a convolutional neural network (CNN) is up to tens of Mbits [2]. Considering hundreds of or thousands of FL clients or even more, the upstream bandwidth consumption could be huge. To achieve optimal learning performance, FL often needs many rounds, where each round involves a client selection process to chose a certain number of clients to participate in the training process. FL can be done both synchronously and asynchronously. The synchronous training requests that the model updates to be carried out within a fixed period so that the CPS can conduct aggregation timely. The synchronous training typically runs more efficiently than the asynchronous one [1] but needs to make sure the local model to be updated within the required time.

Passive optical network (PON) is one of the most promising access technologies to support various residential broadband and mobile services for end users due to its high capacity and energy efficiency [3]. Furthermore, the quality of experience (QoE) for end users with these services can be improved by using emerging machine learning technologies. For services that have privacy concerns, federated learning is a promising approach for QoE improvement. When running synchronized FL training over PONs, a significant challenge is how to efficiently transmit a large amount of traffic under the required synchronization time over PONs to ensure the learning performance. In this regard, our previous work [4] proposed a bandwidth slicing mechanism for PONs to involve a sufficient number of clients within a threshold of synchronization time. The mechanism reserves a certain amount of bandwidth for each FL task, where we found when the involved number of clients increases, the reserved bandwidth becomes the constraint and the learning accuracy may be greatly affected.

In this paper, going beyond [4], we propose two-step aggregation method to facilitate scalable federated learning (SFL) running over PONs, in which the local model parameters are firstly aggregated at optical network units (ONUs) and then further aggregated by the CPS at the central office. In such a way, the required upstream bandwidth for model parameters transmission becomes constant, regardless of the number of clients. Our results reveal that even with limited reserved upstream bandwidth, the number of clients that satisfy synchronization time requirements could be sufficiently large, achieving good learning accuracy.

## 2. Scalable FL over PONs

Figure 1 illustrates the classical and scalable FL over PONs, in which $n$ ONUs are connected with the OLT that is associated with the CPS (i.e., OLT/CPS) at the central office. Each ONU can be co-located with a base station (BS) or an access point (AP), where $m$ mobile devices (e.g., mobile phones) can access wireless networks. These mobile devices (i.e., clients) can work together to train a shared model (e.g., CNN model) for machine learning assisted services (e.g., object recognition) while keeping all the data within the mobile devices. The training is implemented in a round-based manner. At the beginning of the $t^{th}$ round, the CPS selects a certain number of clients to join the FL task and distributes to them a global model, which is parameterized by $w_g^t$. After receiving $w_g^t$, the client_$i,j$ (i.e., the $j^{th}$ client belonging to the $i^{th}$ ONU; $i \in [1,n]$; $j \in [1,m]$) runs local model training with the global model $w_g^t$ and local data ($D_{i,j}$). This results in a local model at the $t^{th}$ round and the parameters can be updated by using stochastic gradient descent (SGD), expressed as $w_{i,j}^{t+1} \leftarrow w_{i,j}^t - \eta \nabla F_{i,j}(w_{i,j}^t, D_{i,j})$, where $\eta$ is the learning rate, and $\nabla F_{i,j}(w_{i,j}^t)$ is the gradient of loss function for the $j^{th}$ client belonging to the $i^{th}$ ONU at the $t^{th}$ round. After completing the local training, if selected the model updates $w_{i,j}^{t+1}$ from the client_i,j are sent to the associated ONU by accessing the BS and further forwarded to the OLT/CPS.

As shown in Fig.1 (a), in the classical case each ONU just forwards the received local model parameters to the OLT/CPS. After gathering updates from all clients, the CPS aggregates local models and updates the global model. Here we employ FedAvg [1]. The global model $w_g^{t+1} \leftarrow \sum_{i=1}^{n}\sum_{j=1}^{m}(k_{i,j}/K)w_{i,j}^t$, where $k_{i,j}$ is the number of data samples for the client_i,j and $K$ is the total number of data samples for all the involved clients. The data volume of all clients' model parameters may be huge, causing a high load on the PON upstream, especially when a large number of clients are involved in the FL task.

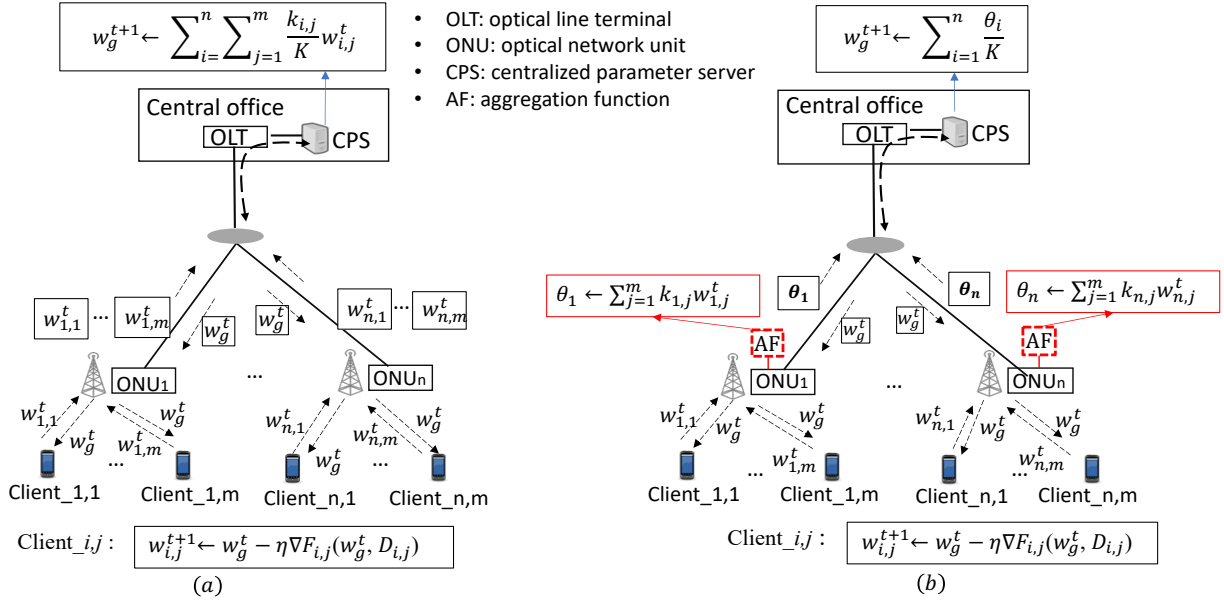

Fig. 1 (a) Classical FL (benchmark) and (b) scalable FL (SFL) over PONs.

The proposed SFL introduces a middle-layer aggregation as shown in Fig. 1(b). Instead of forwarding to the OLT/CPS directly with the local model parameters, each ONU performs one step of model parameter aggregation which then transmits the aggregated model parameters ($\theta$) to the OLT/CPS for further aggregation. An aggregation function (AF) at the ONU just implements simple operations of weighted addition, expressed by $\theta_i \leftarrow \sum_{j=1}^{m} k_{i,j} w_{i,j}^t$, where $i \in [1,n]$ and $j \in [1,m]$. An application protocol interface (API) may be needed to implement this function for ONU management and control. Typically, medium access control (MAC) protocol reserves some bytes for quality of service in control overhead [5, 6], which can be utilized to indicate the model update traffic in Layer 2 and facilitate AF at the ONUs for the SFL. Thanks to the two-step aggregation, the data that needs to be transmitted in the PON upstream can be reduced significantly. After gathering aggregated updates from the ONUs, the CPS updates the global model by $w_g^{t+1} \leftarrow \sum_{i=1}^{n} \theta_i / K$.

### 3. Performance evaluation

A home-made simulator that is developed based on LEAF and Tensorflow [2] is used to evaluate the performance of the proposed SFL over PONs. The dataset of FEMNIST [2] is used for model training, where a CNN model with two 5x5 convolution layers is employed and the FedAvg [1] is chosen for parameter aggregation at the CPS. The data volume for the CNN model update is 26.416 Mbits per client. Hyperparameters (e.g., learning rate, batch size) are set following [2]. Simulations are performed on a workstation with a 10-cores CPU, 64 GB of RAM, 2 TB of storage, and Ubuntu 18.04 LTS. A single FL task is implemented over a time-division multiplexing PON with 16 ONUs, where each ONU/BS can connect up to 20 clients. The distance between the OLT and ONUs is set to 20 km. In each round, the CPS randomly selects clients to be involved in the training.

One-round synchronization time for client_i,j ($T_{i,j}$) consists of global model downloading time ($T_{i,j}^d$), local model training time ($T_{i,j}^r$), and local model uploading time ($T_{i,j}^u$). Therefore, $T_{i,j} = T_{i,j}^d + T_{i,j}^r + T_{i,j}^u$. The global model is broadcast to all the clients via the PON downstream and wireless link. We assume $T_{i,j}^d$ for these clients as a constant, which is set to 2 seconds. $T_{i,j}^r$ depends on the data size, computation resource of clients, and hyperparameters of the learning algorithm. In the simulation, the computation resource and hyperparameters are the same for the clients, thus $T_{i,j}^r$ is mainly related to the size of $D_{i,j}$ and can be obtained during training. In our case, $T_{i,j}^r$ is shown to be between 3 ~20 seconds. In the classical FL, $T_{i,j}^u$ includes the delay on both the wireless networks ($T_{i,j}^w$) and PON upstream ($T_{i,j}^p$). For the wireless part, $T_{i,j}^w$ is distributed in the range of [1, 5] seconds. For the PON upstream, the bandwidth slicing algorithm proposed in [4] is used, and a slice of 100 Mbps is allocated for the FL task. In such a case, $T_{i,j}^p$ is related to the amount of training traffic and can be obtained during simulations. For comparison, model aggregation time ($T_{i,j}^a$) at the $ONU_i$ also needs to be considered in addition to $T_{i,j}^w$ and $T_{i,j}^p$. And, the clients that belong to the same ONU have the same one-round synchronization time. To save training time, the threshold of the one-round synchronization time is set to 25 seconds, and the client_$i,j$ with $T_{i,j} > 25$ seconds (i.e., straggler ) is not able to join the global model updating.

Figure 2(a) shows the upstream bandwidth for one FL task per round in the classical FL (i.e., benchmark) and the proposed SFL. It can be seen that the communication cost saving of the SFL compared to the benchmark is linearly increased as a function of the number of the selected clients ($N$) for training. When $N$=48, such a saving is already 66.7% while it continues to increase to 87.5% when $N$ increases to 128. It indicates that with the number of clients growing, the communication delay in PON of the benchmark increases whereas the proposed FL remains the same. As a result, for the benchmark the number of selected clients that can join the global model updating (i.e., involved clients) may drop due to the required synchronization time threshold. However, it does not happen in the proposed FL. This is shown in Fig. 2(b), where the number of the involved clients for the benchmark is much smaller than the SFL and fluctuates between 1 and 20 for both $N = 48$ and 128. In contrast, the upstream traffic in PON of the proposed SFL is constant and almost all the clients satisfy the synchronization time requirements and can be involved for global model updates, as shown in Fig. 2(b). Since the learning accuracy often depends on the number of involved clients, i.e., the more clients contribute to the training the better learning accuracy can be achieved, it can be expected that the SFL outperforms the benchmark in terms of learning accuracy. The results in Fig. 2(c) clearly reveal that, when N = 128, the learning accuracy for the benchmark is about 0.77, while that for SFL is up to 0.85, leading to an increase of 10%.

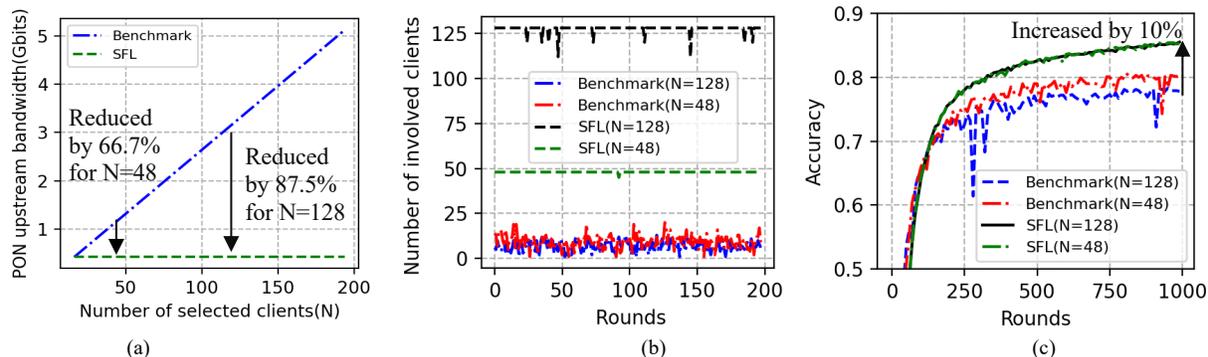

Fig. 2. a) The required upstream bandwidth; b) number of the involved clients; and c) learning accuracy.

### 4. Conclusion

This paper proposes a scalable federated learning (SFL) approach over PONs, which introduces two-step aggregation, namely local model parameters are first aggregated at ONUs and then further aggregated at the OLT by the CPS, to keep the amount of the upstream traffic constant regardless of the number of the involved FL clients. Its advantages on the PON upstream bandwidth saving compared to the classical FL become more obvious when the number of the involved clients increases, leading to higher learning accuracy. Besides, the proposed SFL also potentially improve privacy as the aggregated local models transmitted in PON make the individual local model for each client difficult to interpret.

**Acknowledgment:** This work is supported in part by Swedish Research Council (VR) project 2016-04489 "Go-iData", Swedish Foundation for Strategic Research (SSF), Chalmers ICT-seed grant, and Chalmers CHAIR-seed grant.